\documentclass[aps, prc, reprint, superscriptaddress]{revtex4-1}
\usepackage{amsmath}
\usepackage{amsfonts}
\usepackage{amssymb}
\usepackage{lineno}

\usepackage{graphicx}
\usepackage[caption=false]{subfig}

\usepackage[colorlinks]{hyperref}
\hypersetup{linkcolor=blue, citecolor=blue, filecolor=blue, urlcolor=blue}
\usepackage[all]{hypcap}

\usepackage[utf8]{inputenc}

\newcommand{\IOPP}{Key Laboratory of Quark \& Lepton Physics (MOE) and Institute of Particle Physics, \\
Central China Normal University, Wuhan 430079, China}
\newcommand{\yr}{$N_{{t}} \times N_{{p}} / N_{{d}}^{2}$}
\newcommand{\sNN}{$\sqrt{s_{\mathrm{NN}}}$}

\begin{document}
\title{Light Nuclei Production in Au+Au Collisions at {\sNN} = 5--200 GeV from JAM model}
\author{Hui Liu}
\author{Dingwei Zhang}
\author{Shu He}
\affiliation{\IOPP}
\author{Kai-jia Sun}
\affiliation{Cyclotron Institute and Department of Physics and Astronomy, Texas A\&M University, College Station, Texas 77843, USA}
\author{Ning Yu}
\affiliation{School of Physics \& Electronic Engineering, Xinyang Normal University, Xinyang 464000, China}
\author{Xiaofeng Luo}
\email{xfluo@mail.ccnu.edu.cn}
\affiliation{\IOPP}
\footnotetext[1]{You can add acknowledgements here.}
\footnotetext[2]{You can add acknowledgements here.}

\begin{abstract}

Light nuclei production is sensitive to the baryon density fluctuations and can be used to probe the QCD phase transition in relativistic heavy-ion collisions. In this work, we studied the production of proton, deuteron, triton in central Au+Au collisions at $\sqrt{s_{\mathrm{NN}}}$  =  5, 7.7, 11.5, 14.5, 19.6, 27, 39, 54.4, 62.4 and 200 GeV from a transport model (JAM). Based on the coalescence production of light nuclei,  we calculated the energy dependence of rapidity density $dN/dy$ and particle ratios ($d/p$, $t/p$, and $t/d$). More importantly, the yield ratio {\yr}, which is sensitive to the neutron density fluctuations, shows a flat energy dependence and cannot describe the non-monotonic trend observed by the STAR experiment. Based on the nucleon coalescence, this work can provide constraint and reference to search for the QCD critical point and/or first order phase transition with light nuclei production in future heavy-ion collision experiments. \end{abstract}

\maketitle

\section{INTRODUCTION}
Understanding the Quantum Chromodynamics (QCD) phase diagram of strongly interacting matter is of fundamental importance in nuclear physics. The QCD phase diagram can be displayed in the two dimensional phase diagram of temperature ($T$) versus baryon chemical potential ($\mu_{\mathrm{B}}$). Lattice QCD calculations show that 
the transition from hadronic phase to quark-gluon plasma (QGP) is smooth crossover at small values of $\mu_{\mathrm{B}}$~\cite{Aoki:2006we}. While at finite $\mu_{\mathrm{B}}$, it is of first order phase transition based on QCD model calculations~\cite{Endrodi:2011gv}. If those predictions are true, by definition, there should be a QCD critical point as the  end point of the first order phase boundary. However, there is still large uncertainties in determining the location and even the existence of the QCD critical point from theoretical side. Experimentally, relativistic heavy-ion collisions can provide us a useful and controllable way to explore the QCD phase structure, especially on finding the QCD critical point~\cite{Gupta:2011wh,Luo:2015doi,Luo:2017faz,Bzdak:2019pkr,Aggarwal:2010wy,Adamczyk:2013dal,Adamczyk:2014fia,Adamczyk:2017wsl,Adam:2020unf}. This is one of the main physics motivation of the Beam Energy Scan program (BES-I, 2010-2014 \& BES-II: 2019-2021) at Relativistic Heavy-ion Collider (RHIC)~\cite{Aggarwal:2010cw}.

\begin{figure*}[htb]\label{fig:SPECTRA}
\centering
    \subfloat{\label{fig:pt_p}
        \includegraphics[width=0.65\columnwidth]{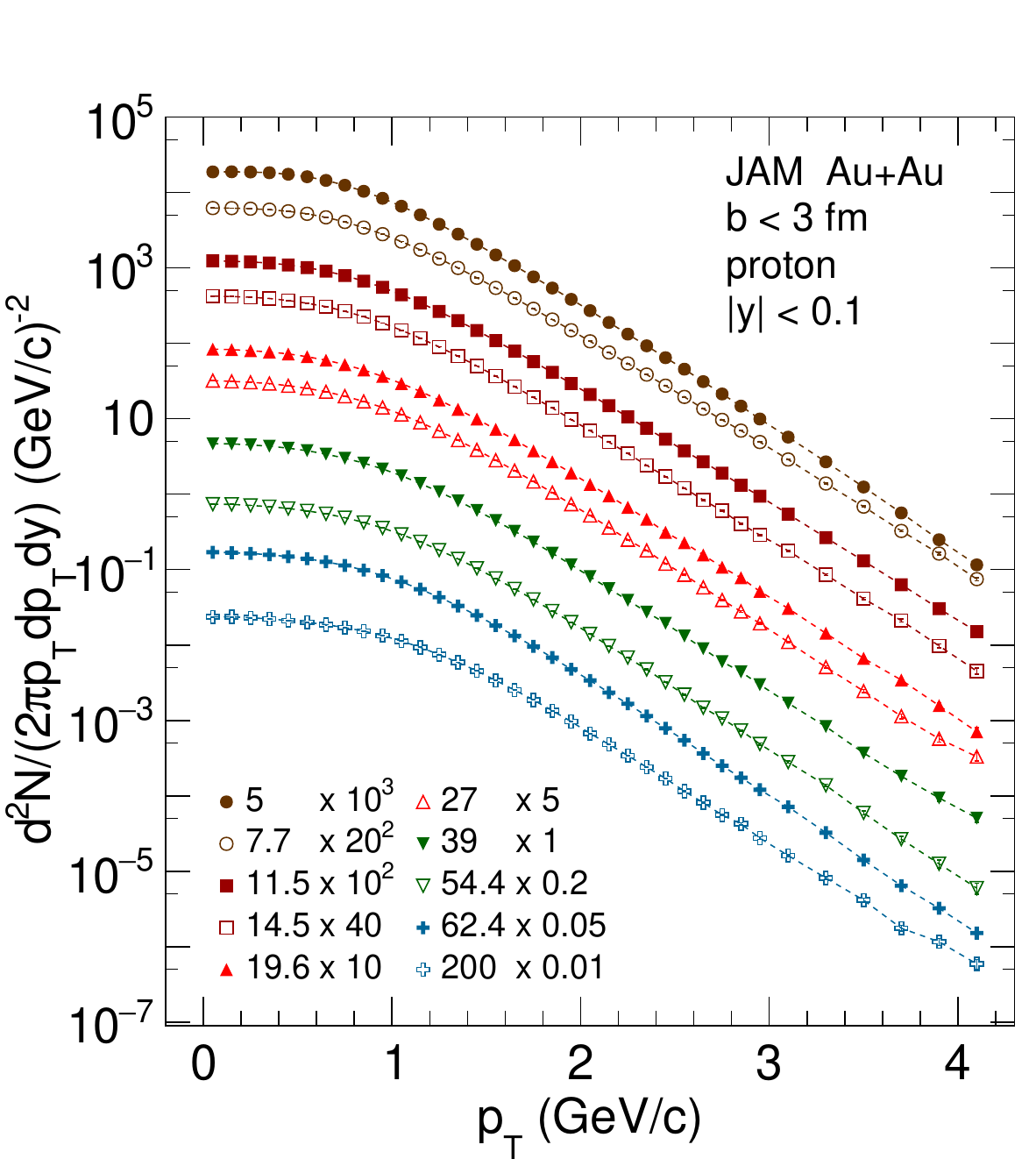}
    }
    \subfloat{\label{fig:pt_d}
        \includegraphics[width=0.65\columnwidth]{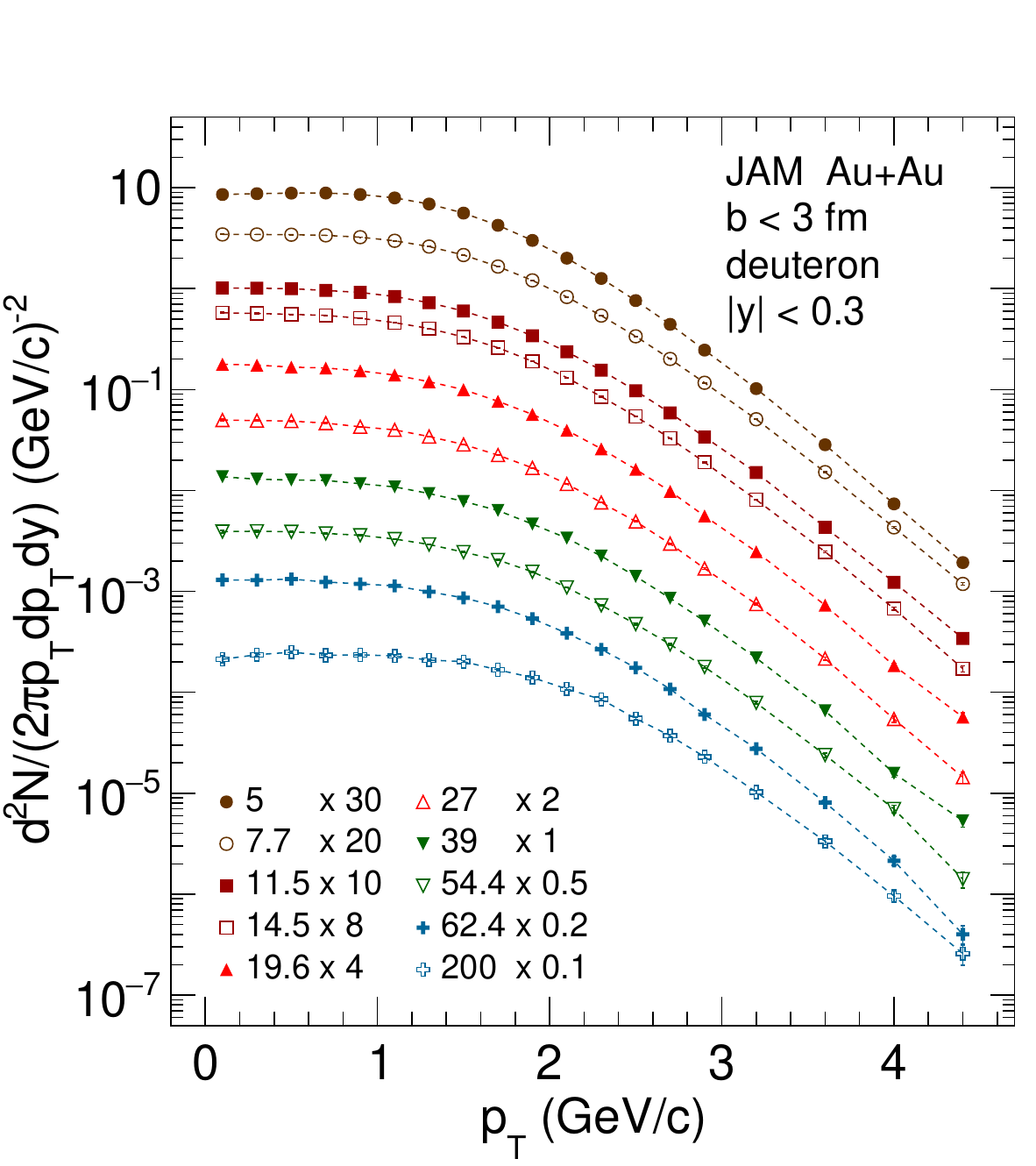}
    }
    \subfloat{\label{fig:pt_t}
        \includegraphics[width=0.65\columnwidth]{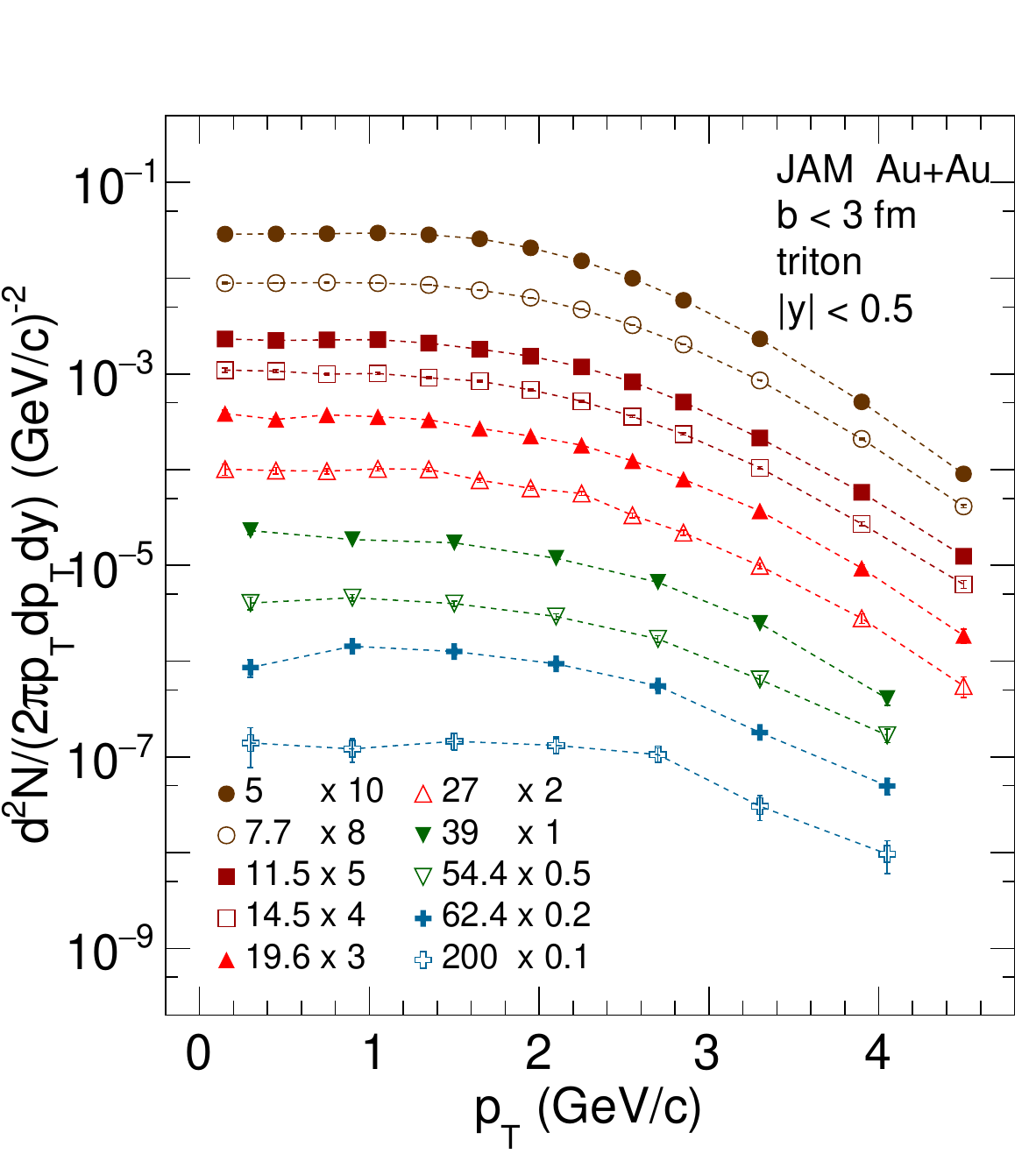}
    }
    \caption{Transverse momentum spectra for proton, deuteron and triton in most central ($b<3 fm$) Au + Au collisions at $\sqrt{s_{\mathrm{NN}}}$  =  5, 7.7, 11.5, 14.5, 19.6, 27, 39, 54.4, 62.4 and 200 GeV within JAM model. For illustration purpose, different energies are scaled by different factors. The rapidity cuts for proton ($|y|<0.1$), deuteron ($|y|<0.3$) and triton ($|y|<0.5$) are chosen to be the same as the ones used in the STAR data analysis.}
\end{figure*}
Light nuclei, such as deuteron and triton, are loosely bounded objects with small binding energies ($d$ with 2.2 MeV and $t$ with 8.4 MeV). Those are formed via coalescence of nucleons in a very restricted phase-space volume~\cite{Abelev:2010rv,Agakishiev:2011ib,Andronic:2017pug,Braun-Munzinger:2018hat}. Because of the small binding energies, their existence at high temperature environment  created in heavy-ion collisions is called "snowball in hell"~\cite{Andronic:2017pug}, which seems counter intuitive. For example, at LHC energies, the yield of deuteron, triton and even hypertriton in Pb+Pb collisions at \sNN\ = 2.76 TeV measured by ALICE experiment~\cite{Adam:2015vda} can be well described by thermal model, which is hard to understand. A recent hybrid model (hydrodynamics + hadronic afterburner) study of deuteron production at LHC~\cite{Oliinychenko:2018ugs} energy provide a possible explanation to the "snowball in hell" phenomena and the successful thermal description of ALICE data: the initially thermal produced light nuclei are dynamically destroyed and re-generated via detail balanced hadronic re-scattering process in heavy-ion collisions, which keeps the light nuclei yield almost unchanged. This indicates the hadronic re-scattering stage play a very important role for light nuclei production in heavy-ion collisions. This result is also shown in an early work, in Ref.~\cite{Oh:2009gx}, they study deuteron production by a hadron transport model and compared with those measured by STAR Collaborations for Au + Au collisions at $\sqrt{s_{\mathrm{NN}}}$= 200 GeV. At RHIC BES energies, the STAR experiment has collected the data of Au+Au collisions at $\sqrt{s_{\mathrm{NN}}}$= 7.7, 11.5, 14.5, 19.6, 27, 39, 54.4 62.4 and 200 GeV and measured the light nuclei production (deuteron and triton)~\cite{Adam:2019wnb, Zhang:2019wun,Zhang:2020ewj}. It was found that thermal model can describe the deuteron yield well, but overestimate the triton yield in central Au+Au collisions at \sNN\ = 7.7--200 GeV measured by STAR experiment. This could be related to the different production mechanisms of light nuclei at different energies, which still remains an open question.

On the other hand, due to increasing of the correlation length and formation of instability spinodal domain, both of the critical fluctuations and first order phase transition can induce large baryon density fluctuations. It is predicted that the production of light nuclei is sensitive to the baryon density fluctuations and thus can be used to probe the QCD phase transition in heavy-ion collisions~\cite{Sun:2017xrx,Sun:2018jhg,Shuryak:2019ikv,Yu:2018kvh,Shao:2019xpj}. For instance, the neutron density fluctuation ($\Delta n=\left\langle(\delta n)^{2}\right\rangle /\langle n\rangle^{2}$) can be extracted from the yield ratio of proton, deuteron and triton {\yr}. Interestingly, it was observed that the yield ratio and the extracted neutron density fluctuation ($\Delta n$) in central Au+Au collisions measured by STAR experiment shows clear non-monotonic energy dependence,  with a peak around 20 GeV~\cite{Zhang:2019wun,Zhang:2020ewj}. However, without dynamical modeling of the QCD phase transition in heavy-ion collisions, it is difficult to give a definitive conclusion about the signature of the QCD critical point and/or first-order phase transition.

In this paper, we studied the production of deuteron and triton in most central ($b<3fm$) Au+Au collisions at $\sqrt{s_{\mathrm{NN}}}$ = 5, 7.7, 11.5, 14.5, 19.6, 27, 39, 54.4, 62.4 and 200 GeV from JAM model. Our paper are organized as following: In Sec.~\ref{section:JAM model}, we give a brief introduction to the JAM model. In Sec.~\ref{section:neutron density fluctuation}, we show the relations between neutron density fluctuation and the yield ratio in relativistic heavy-ion collisions. In Sec.~\ref{section:Result}, we present the mid-rapidity transverse momentum spectra for proton, deuteron and triton. Furthermore, we show the energy dependence of particle yield $dN/dy$, the particle ratios ($d/p$, $t/p$ and $t/d$), and the yield ratio {\yr}. Finally, the summary will be given in section~\ref{section:summary}.

\section{JAM MODEL}\label{section:JAM model}
In relativistic heavy-ion collisions, the whole process from first $NN$ collision stage to the final state interaction among produced particles is very complicated and  involves a lot of dynamic evolution. To explore these evolutionary processes, many microscopic hadronic transport models are used to simulate the relativistic heavy-ion collisions, such as RQMD~\cite{Sorge:1995dp, Sorge:1997nv}, UrQMD~\cite{Bass:1998ca, Bleicher:1999xi}, ARC~\cite{Kahana:1997aq}, ART~\cite{Li:1997pu} and AMPT~\cite{Lin:2004en}. JAM model~\cite{Nara:2019crj}(Jet AA Microscopic Transportation Model) has been developed based on the resonance and string degrees of freedom. In JAM model, particles are produced via resonance or excitation of strings and their decays.  Hadrons and their excited states have explicit space and time evolution trajectories by the cascade method. Inelastic hadron-hadron collisions are modeled with resonance at low energy, string picture at intermediate energy and hard parton-parton scattering at high energy. The nuclear mean-field is implemented based on the simplified version of the relativistic quantum molecular dynamics (RQMD) approach. It is a skyrme-type density dependent and Lorentzian-type momentum dependent scalar mean-field potential~\cite{Isse:2005nk}. More details can be found in Refs.~\cite{Nara:1999dz, Nara:2016phs,Nara:2017qcg}. 
In this work, we focus on the production of light nuclei in Au+Au collisions at $\sqrt{s_{\mathrm{NN}}}$ = 5--200 GeV. The JAM model doesn't directly generate the light nuclei from the model itself. Instead, the light nuclei is produced with an afterburner code via coalescence of nucleons with the phase space obtained from the JAM model. The coalescence conditions are controlled by two parameters, which are the relative distance ($\Delta \mathrm{R}$) and relative momentum ($\Delta \mathrm{P}$)~\cite{Sombun:2018yqh} in the two-body center-of-mass frame. 
For coalescence process, if the relative distance and momentum of any two nucleons in a light nuclei are less than the given parameters ($\mathrm{R}_{0}$, $\mathrm{P}_{0}$)~\cite{Li:2016mqd, Nara:2015ivd}, the light nuclei is   considered to be formed. Based on the charge rms radius of wave function for deuteron and triton~\cite{Bellini:2018epz}, we fixed the coalescence parameters of deuteron and triton $\Delta \mathrm{R}$ = 4 and 3.4 $fm$, respectively.  The $\Delta \mathrm{P}$ of deuteron and triton are chosen to be the same values $\Delta \mathrm{P}$ = 0.3 GeV/c, with which the STAR measured light nuclei yield ratio {\yr} in central Au+Au collisions at $\sqrt{s_{\mathrm{NN}}}$ = 200 GeV can be  reproduced. 

\begin{figure}[h]\label{fig:PARTICLE YIELD}
\centering
    \subfloat{\label{fig:dndy}
        \includegraphics[width=1\columnwidth ]{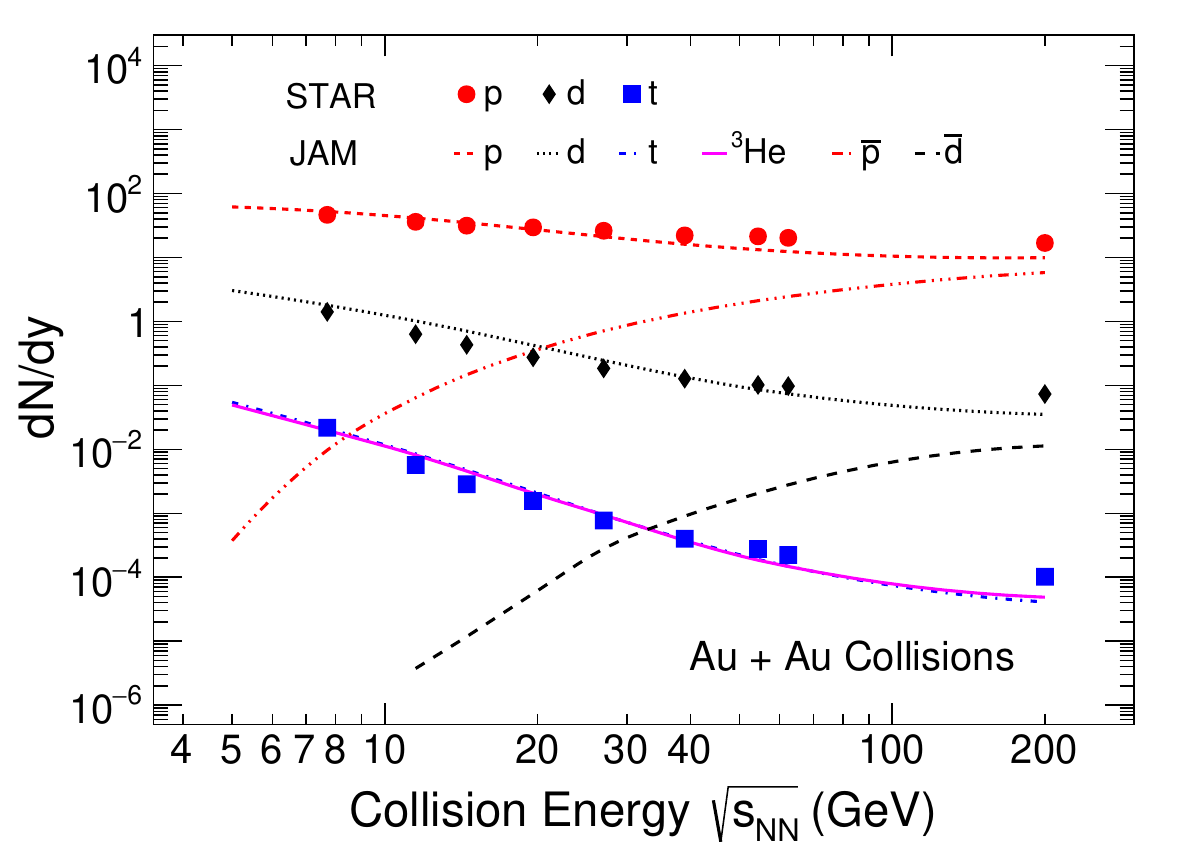}
    }
    \caption{Energy dependence of mid-rapidity $dN/dy$ of $p$, $d$, $t$, $^3\mathrm{He}$ and $\bar{p}$, $\bar{d}$ in most central  ($b<$ 3 fm) Au + Au collisions from JAM model (lines).  The experimental results of 
    $p$, $d$ and $t$ from STAR experiment are also presented as markers for comparison. }
\end{figure}
%
\begin{figure}[h]\label{fig:PARTICLE RATIO}
\centering
    \subfloat{\label{fig:particle_ratio}
        \includegraphics[width=1\columnwidth]{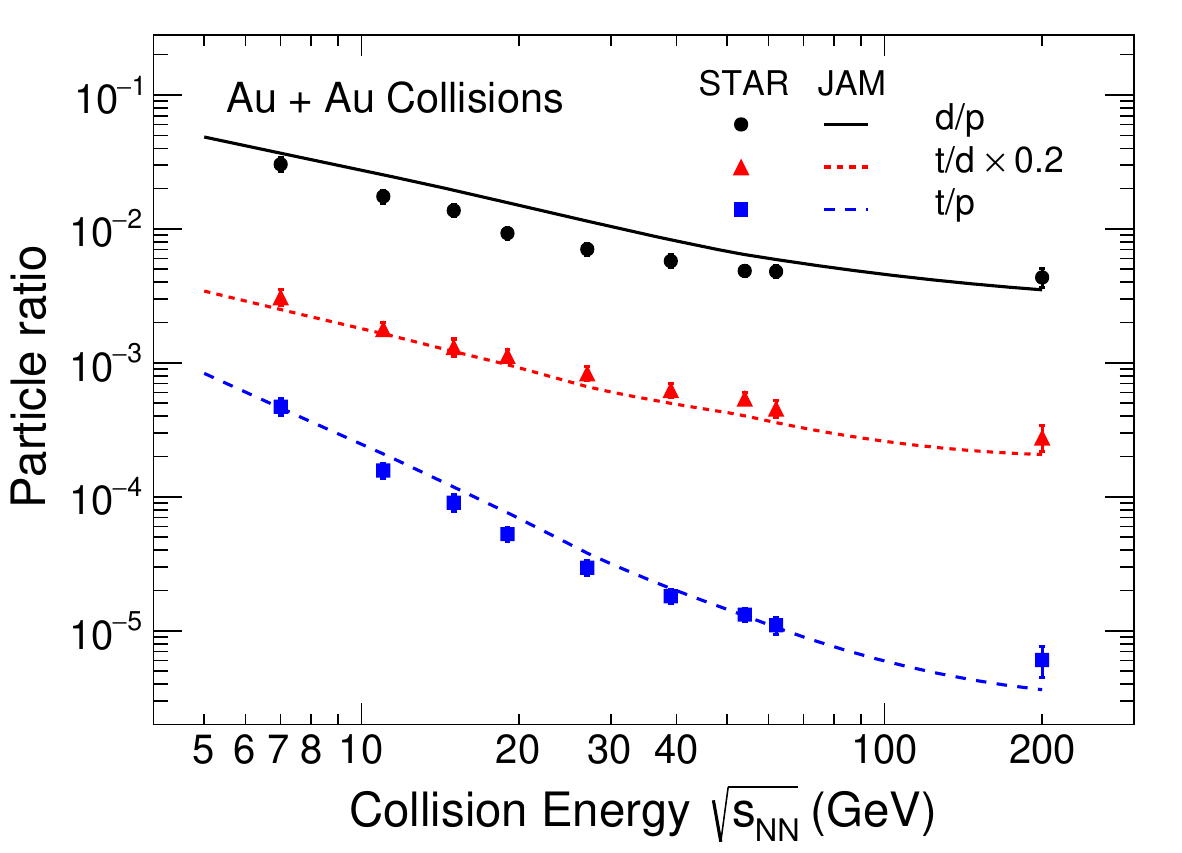}
    }
    \caption{Energy dependence of particle ratio $d/p$ (black dots), $t/p$ (blue squares), $t/d$ (red triangles) in central Au+Au collisions at {\sNN} =  7.7--200 GeV measured by STAR experiment.  The error bars of the STAR data are the quadrature sum of statistical and systematic uncertainties. The lines represent the results from JAM model calculations. }
\end{figure}
\section{Relation between NEUTRON DENSITY FLUCTUATION and Yield Ratio of Light Nuclei}\label{section:neutron density fluctuation}
Based on the coalescence model~\cite{Sun:2017xrx},  the light nuclei is formed via nucleon coalescence and  the neutron density fluctuation ($\Delta n=\left\langle(\delta n)^{2}\right\rangle /\langle n\rangle^{2}$) at kinetic freeze out can be encoded in the yield ratio of light nuclei (\yr). 
In the following, we will briefly introduce the relations between the neutron density fluctuations and light nuclei yield ratio {\yr}, which is based on the Refs.~\cite{Sun:2017xrx, Sun:2018mqq}. 

In the coalescence model, the number of deuteron and triton can be expressed approximately as~\cite{Sun:2017xrx, Sun:2018mqq}. 
\begin{equation}\label{eq:deuteron number 1}
\begin{aligned}
    N_{{d}}=\frac{3}{2^{1 / 2}}\left(\frac{2 \pi}{m_{0} T_{\mathrm{eff}}}\right)^{3 / 2} \frac{N_{p} N_{n}}{V}
\end{aligned}
\end{equation}
\begin{equation}\label{eq:triton number 1}
\begin{aligned}
    N_{{t}}=\frac{3^{3 / 2}}{4}\left(\frac{2 \pi}{m_{0} T_{\mathrm{eff}}}\right)^{3} \frac{N_{p} N_{n}^{2}}{V^{2}}
\end{aligned}
\end{equation}
where $N_{p}$ and $N_{n}$ are the number of protons and neutrons, respectively. $V$ is system volume and $T_\mathrm{eff}$ is the effective local temperature at kinetic freeze-out. In this coalescence picture, uniform distributions of nucleons in space are assumed. One can also neglect the mass difference between neutron and proton and set $m_{p}=m_{n}=m_{0}$. 
If we ignore the binding energies of deuteron and triton, the results obtained in Eq.(\ref{eq:deuteron number 1}) and Eq.(\ref{eq:triton number 1}) are consistent with thermal model with the assumption that nucleon, deuteron and triton are in fully thermal and chemical equilibrium at kinetic freeze out~\cite{BraunMunzinger:2007zz, Andronic:2010qu, Steinheimer:2012tb,Oliinychenko:2020ply}.

Based on Eq.(\ref{eq:deuteron number 1}) and Eq.(\ref{eq:triton number 1}), we introduce neutron density fluctuations, i.e $n(\vec{r})=\frac{1}{V} \int n(\vec{r}) \mathrm{d} \vec{r}+\delta n(\vec{r})=\langle n\rangle+\delta n(\vec{r})$, here $\delta n(\vec{r})=0$ with uniform distribution, we can rewrite Eq.~\ref{eq:deuteron number 1} and Eq.~\ref{eq:triton number 1} as
\begin{equation}\label{eq:deuteron number 2}
\begin{aligned}
    N_{{d}}=\frac{3}{2^{1 / 2}}\left(\frac{2 \pi}{m_{0} T_{\mathrm{eff}}}\right)^{3 / 2} N_{p}\langle n\rangle(1+\alpha \Delta n),
\end{aligned}
\end{equation}
\begin{equation}\label{eq:triton number 2}
\begin{aligned}
    N_{{t}}=\frac{3^{3 / 2}}{4}\left(\frac{2 \pi}{m_{0} T_{\mathrm{eff}}}\right)^{3} N_{p}\langle n\rangle^{2}[1+(1+2 \alpha) \Delta n],
\end{aligned}
\end{equation}
where $\alpha$ is correlation coefficient between neutron and proton number density. If we assume the correlation between density of protons and neutrons are small ($\alpha \approx 0$), then we have : 
\begin{equation}\label{eq:yield ratio 1}
\begin{aligned}
      \frac{N_{{t}} \times N_{{p}}}{N_{{d}}^{2}}={g}(1+\Delta n)
\end{aligned}
\end{equation}
where $g = \frac{1}{2\sqrt{3}} \approx 0.29$. In Eq.~\ref{eq:yield ratio 1}, one can see that the neutron density fluctuation can be probed by measuring the yield ratio of light nuclei in heavy-ion collisions.
Thus, the light nuclei production can provide us an useful tool to study the QCD phase transition.

\begin{figure*}\label{fig:YIELD RATIO}
\centering
    \subfloat{\label{fig:yield_ratio}
        \includegraphics[width=1.3\columnwidth]{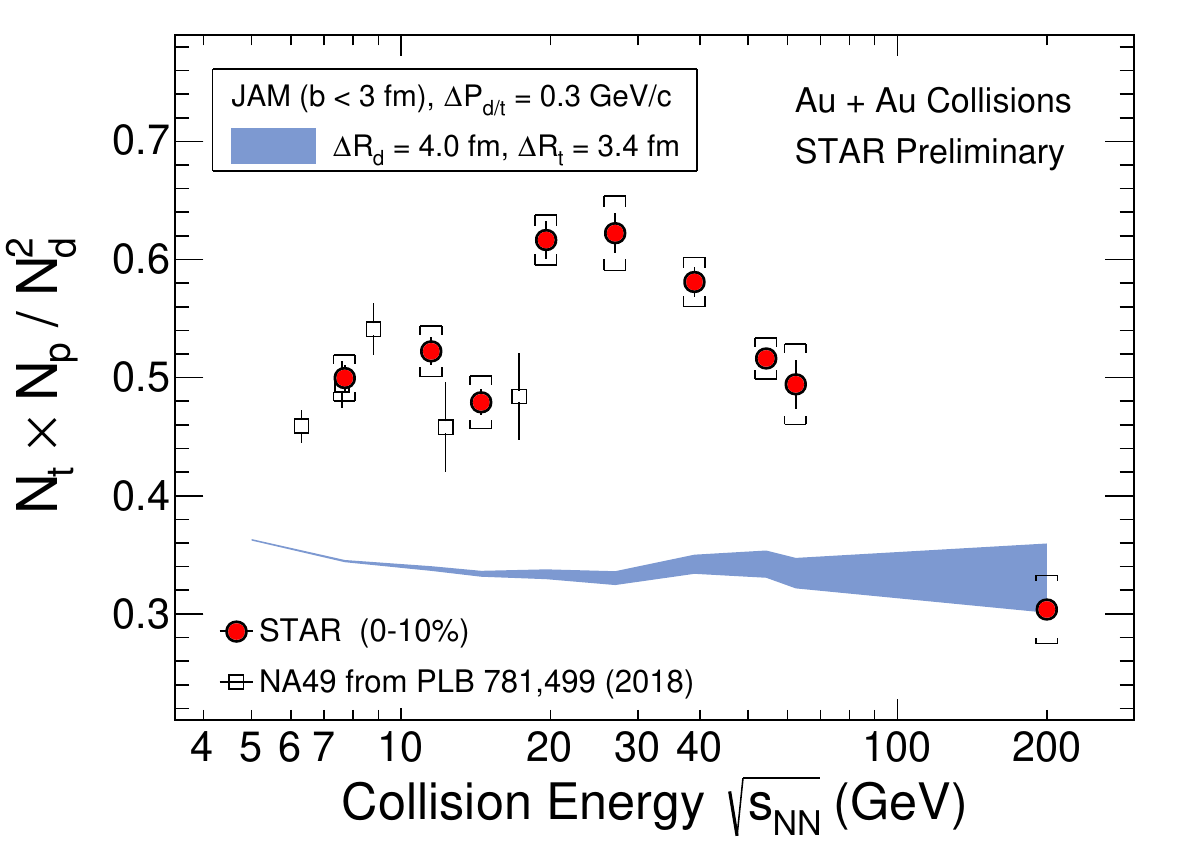}
    }
    \caption{Energy dependence of the yield ratio of light nuclei {\yr} in central Au+Au collisions from JAM model (blue band) and its comparison with the experimental data measured by the STAR (red solid circles)~\cite{Zhang:2019wun,Zhang:2020ewj} and NA49 (black empty square)~\cite{Sun:2018jhg,Anticic:2016ckv} experiment. The results from NA49 experiment include the data of central Pb+Pb collisions at  {\sNN} (GeV) = 6.3 (0-7$\%$), 7.6 (0-7$\%$), 8.8 (0-7$\%$), 12.3 (0-7$\%$), and 17.3 (0-12$\%$). The vertical bars and caps of the STAR data are the statistical and systematic uncertainties, respectively.  The error bars of the NA49 data are the quadrature sum of statistical and systematic uncertainties.}
\end{figure*}
%
\section{RESULTS}\label{section:Result}
In this section, we present the transverse momentum spectra, $dN/dy$ and yield ratios in most central Au+Au collisions at \sNN\ = 5, 7.7, 11.5, 14.5, 19.6, 27, 39, 54.4, 62.4 and 200 GeV from JAM model. The minimum bias collisions were chosen and the impact parameter is set to be less than 3 $fm$. 
\subsection{Transverse Momentum Spectra and $dN/dy$}
Figure~\ref{fig:SPECTRA} shows the transverse momentum spectra for proton, deuteron and triton in most central ($b<3fm$) Au + Au collisions at $\sqrt{s_{\mathrm{NN}}}$  =  5, 7.7, 11.5, 14.5, 19.6, 27, 39, 54.4, 62.4 and 200 GeV from JAM model. For illustration purpose, we scaled the spectra with different constant factors for various energies. Fig.~\ref{fig:PARTICLE YIELD} shows the energy dependence of $dN/dy$ of $p$, $d$, $t$, $^3\mathrm{He}$ and $\bar{p}$, $\bar{d}$. Comparing the results of particle with anti-particle~\cite{Chen:2018tnh, Adamczyk:2017iwn,Oliinychenko:2018ugs}, we find the particle yields decrease with the increasing energy, while the yield of anti-particles increase. This is due to the interplay of baryon stopping and pair production of nucleons at different energies: baryon stopping dominated at low energies, while the pair production dominated at high energies~\cite{Cleymans:2005zx, Luo:2017faz}, which make the yields of particle and anti-particle get closer at higher energies. We also found the yields of $t$ and $^3\mathrm{He}$ are consistent within uncertainties, as they are both coalesced from three nucleons.

\subsection{Light Nuclei Yield Ratios}
Using the integrated yields, one can obtain the particle ratios ($d/p$, $t/p$, $t/d$) as a function of collision energy. As shown in Fig.~\ref{fig:PARTICLE RATIO}, the lines calculated by JAM model are compared to the preliminary results from central Au+Au collisions at RHIC BES energies ( {\sNN} = 7.7 to 200 GeV) measured by the STAR experiment~\cite{Adam:2019wnb, Zhang:2019wun, Zhang:2020ewj}. 
Generally, the energy dependence trends of particle ratios, $d/p$, $t/p$, $t/d$, can be qualitatively described by the results from JAM model. It was observed that the energy dependence of $d/p$ and $t/p$ ratios from JAM model  are slightly larger than the experimental data measured by STAR. For $t/d$ ratio, the result from JAM model is in good agreement with the STAR data. In addition, those ratios decrease with increasing of collision energy, which can be explained by the lower baryon density at higher energy. We want to emphasize that since the production mechanism of light nuclei at different energies are still not well understood yet, we are not expecting the particle ratio $d/p$, $t/p$ and $t/d$ over broad energy range can be simultaneously well described by the model with the simple coalescence method. 

As discussed in section III, the yield ratio {\yr} is related to the neutron density fluctuation and can serve as a sensitive observable to probe the QCD phase transition in heavy-ion collisions. The STAR experiment has measured the deuteron and triton production in Au+Au collisions at {\sNN}= 7.7 to 200 GeV~\cite{Adam:2019wnb, Zhang:2019wun,Zhang:2020ewj}. Fig.~\ref{fig:YIELD RATIO} shows the energy dependence of the yield ratio {\yr} in central Au+Au collisions from JAM model and its comparison with the experimental data measured by STAR experiments. The experimental data from STAR shows a clear non-monotonic energy dependence with a peak around 20 GeV. At energies below 20 GeV, the results from STAR~\cite{Zhang:2019wun,Zhang:2020ewj} and NA49~\cite{Sun:2018jhg,Anticic:2016ckv}  experiment are consistent. The yield ratio calculated from the JAM model is much smaller than the experimental data and show a flat energy dependence. We note that it is possible to have a better description of $d/p$ ratio or $t/p$ ratio by tuning the coalescence parameter, but this only leads to a overall shift of {\yr} ratio and the flat energy dependence will remain similar. Non-monotonic energy dependence of light nuclei yield ratio could be related to the large baryon density fluctuation near the critical point or first-order phase transition. To provide a definite physics conclusion on this non-monotonic behavior, we still need dynamical modeling of the heavy-ion collisions with more realistic equation of state. Our transport model study, which is without the physics of QCD phase transition, can provide constraint and reference for future QCD critical point search in heavy-ion collisions.

\section{SUMMARY}\label{section:summary}
We presented the light nuclei production in central Au+Au collisions at $\sqrt{s_\mathrm{NN}}$ = 5, 7.7, 11.5, 14.5, 19.6, 27, 39, 54.4, 62.4 and 200 GeV within JAM model. The light nuclei is produced via the coalescence of final state nucleons in the phase space from the model. We showed the transverse momentum spectra for proton, deuteron and triton, and the energy dependence of yields and their ratios. It was observed that the particle ratios $d/p$ and $t/p$ from JAM model are slightly larger than the experimental data from STAR experiment, while the values of triton to deuteron ratio ($t/d$) is in good agreement with the STAR data. On the other hand, it is predicted that the light nuclei yield ratio {\yr} can be used to probe the neutron density fluctuation in heavy-ion collisions. We analyzed the energy dependence of light nuclei yield ratio {\yr} in central Au+Au collisions from JAM model and compared it with the experimental data measured by STAR experiment. We found that the values of yield ratio {\yr} from JAM model are much smaller than the experimental data and shows a flat energy dependence, which cannot describe
the non-monotonic energy dependence trend observed by the STAR experiment. One should keep in mind that there is only hadronic degree of freedom in JAM model and the physics of QCD phase transition are not implemented.  Based on the nucleon coalescence, this model study can provide constraint and reference to search for the QCD critical point and/or the first-order phase transition with light nuclei production in the future heavy-ion collision experiments. 
\section{Acknowledgement}
We thank Liewen Chen,  Che Ming Ko, Yugang Ma, Dmytro Oliinychenko, Lijuan Ruan, Huichao Song, Nu Xu, Zhangbu Xu, Pengfei Zhuang and Wenbin Zhao for stimulating discussion. This work is supported by the National Key Research and Development Program of China (2018YFE0205201),  the National Natural Science Foundation of China (No.11828501, 11575069, 11890711 and 11861131009).  

\end{document}